%% file: v4.2_CoulombPhases_arXiv.tex
\documentclass[prd,aastex,amsmath,amssymb,twocolumn,floatfix,aps,showpacs]{revtex4-2} 

\usepackage{graphicx}% Include figure files
\usepackage{amsmath}
\usepackage{amssymb}
\usepackage{textcomp}
\usepackage{xcolor}
\definecolor{myblue}{rgb}{0.0, 0.0, 0.6}
\usepackage{hyperref}
\hypersetup{
  colorlinks = true,
  citecolor  = myblue,
  linkcolor  = myblue,
  urlcolor   = myblue
}
\begin{document}
\title{
  Coulomb phase corrections to the transverse analyzing power $A_\text{N}(t)$ in high energy forward proton-proton scattering}%

\author{A.~A.~Poblaguev}\email{poblaguev@bnl.gov}

\affiliation{Brookhaven National Laboratory, Upton, New York 11973, USA}

\date{May 20, 2022}% It is always \today, today,

\begin{abstract}
  Study of polarized proton-proton elastic scattering in the Coulomb-nuclear interference region allows one to measure the forward hadronic single spin-flip amplitude including its phase. However, in a precision experimental data analysis, a phase-shift correction $\delta_C$ due to the long distance Coulomb interaction should be taken into account. For unpolarized scattering,  $\delta_C$ is commonly considered as well established. Here, we evaluate the Coulomb phase shifts for the forward elastic proton-proton single spin-flip electromagnetic and hadronic amplitudes. Only a small discrepancy between the spin-flip and nonflip phases was found which can be neglected in the high energy forward elastic $\mathit{pp}$ studies involving transverse spin. Nonetheless, the effective alteration of the hadronic spin-flip amplitude by the long-distance electromagnetic corrections can be essential for interpretation of the experimental results.
\end{abstract}

\maketitle

\section{Introduction}

Since electromagnetic amplitude can substantially contribute to the elastic forward proton-proton ($\mathit{pp}$) scattering at high energies, experimental study of the Coulomb-nuclear interference (CNI) in the ${pp}$ scattering allows one to reveal the hadronic amplitude structure. For the unpolarized scattering, the CNI ${pp}$ amplitude can be approximated as\,\cite{Block:1984ru}
\begin{equation}
  \phi_{pp}^\text{CNI}(s,t) = \text{Im}\,\phi(s,0)\left[
    (i\!+\!\rho)e^{Bt/2}\!+\!\frac{t_c}{t}e^{i\delta_C+\widetilde{B}t/2}
  \right]\!.
    \label{eq:CNI}
\end{equation}
Here, $\phi(s,t)$ stands for the hadronic amplitude and the electromagnetic component is identified by $t_c/t$ term, where $t_c\!=\!-8\pi\alpha/\sigma_\text{tot}(s)$, $\alpha$ is the fine structure constant and $\sigma_\text{tot}$ is the total $pp$ cross section.
Generally, $\phi_{pp}^\text{CNI}$ and the parameters used are functions of total energy squared $s$ and momentum transfer squared $t$.
  
In this paper, numerical estimates will be done for a 100\,GeV proton beam (typical for the Relativistic Heavy Ion Collider and the future Electron Ion Collider) scattering  off a fixed proton target. Therefore, %
$\rho\!=\!-0.079$\,\cite{Fagundes:2017iwb},  
$\sigma_\text{tot}\!=\!39.2\,\text{mb}$\,\cite{Fagundes:2017iwb},  
$t_c\!=\!-1.86\!\times\!10^{-3}\,\text{GeV}^2$,
and $B\!=\!11.2\,\text{GeV}^{-2}$\,\cite{Bartenev:1973jz,*Bartenev:1973kk}.
The electromagnetic form factor is expressed,
\begin{equation}
  \widetilde{B} = 2r_E^2/3 = 12.1\,\text{GeV}^{-2},
  \label{eq:Bem}
\end{equation}
via rms charge radius of a proton, $r_E\!=\!0.841\,\text{fm}$\,\cite{Zyla:2020zbs}.

For the unpolarized scattering, a theoretical understanding of the Coulomb phase shift $\delta_C(t)$ was developed in many works, particularly in \cite{Akhiezer:1945,*Bethe:1958zz,*West:1968du,*Cahn:1982nr}. Following Ref.\,\cite{Kopeliovich:2000ez} and neglecting terms $\sim\!t^2\ln{t}$,
\begin{align}
  \delta_C(t)/\alpha &= -\ln{\left[(B+\widetilde{B})|t|/2\right]} -\gamma
    \label{eq:dC} \\
    &+  \frac{\widetilde{B}t}{2}\,
    \left( \ln{\frac{\widetilde{B}|t|}{2}}\!+\!\gamma\!+\!\ln{2}\!-\!1\right)
    -\frac{B}{B\!+\!\widetilde{B}}\frac{Bt}{2},
    \label{eq:dC_}
\end{align}
where $\gamma\!=\!0.5772$ is Euler's constant. The leading order approximation (\ref{eq:dC}) is commonly used in experimental data analysis for many years. The next to leading order corrections (\ref{eq:dC_}) can be disregarded in this paper.

It was shown in Ref.\,\cite{Buttimore:1978ry} that the Coulomb phase should be independent of the helicity structure of the experimentally measured scattering amplitudes though subsequent determination of the pure hadronic amplitudes may involve order-$\alpha$ corrections resulting from the spin of the particles involved.

Recently, long-distance electromagnetic corrections, including the absorption, to the spin-flip amplitudes were derived\,\cite{Kopeliovich:2021rdd} in the eikonal model. However, the results were presented as Fourier integrals (which were calculated numerically) and, thus, cannot be implemented, in a simple way, to an experimental data analysis. Also, it was noted\,\cite{Kopeliovich:2021rdd} that the corrections to the spin-flip Coulomb phase ``are so large, that hardly can be treated as a phase shift'', which may be understood as a suggestion to obsolete a commonly used expression for the analyzing power $A_\text{N}(t)$ given in Eq.\,(\ref{eq:AN}) and, consequently, to reanalyze all previous measurements of $A_\text{N}(t)$.

Here, we found compact algebraic approximations for the long-distance electromagnetic corrections, separately, to the spin-flip Coulomb phase and to the hadronic spin-flip amplitude parameter $r_5$ [Eq.\,(\ref{eq:r5})]. It was shown that the difference between the spin-flip and nonflip Coulomb phases is small. Although, the effective correction to $\text{Re}\,r_5$ was found to be essential for the experimental accuracy already achieved\,\cite{Poblaguev:2019saw}, it can be applied directly to the value of $r_5$ obtained in experimental data fit. The corresponding changes in the Regge fit\,\cite{Poblaguev:2019saw} of the measured values of $r_5(s)$ will be discussed.

\section{High energy forward elastic $\boldsymbol{pp}$ analyzing power
\label{sec:AN}}

For elastic scattering ${p^\uparrow}{p}$ of a vertically polarized proton beam off a proton target, the analyzing power $A_N$ is defined by the interference of the nonflip ({\em nf}) and spin-flip ({\em sf}) helicity amplitudes\,\cite{Kopeliovich:1974ee,Buttimore:1978ry,Buttimore:1998rj}
\begin{equation}
  A_N = \frac{2\,\text{Im}\left[%
      \widetilde{\phi}_\textit{sf}\phi_\textit{nf}^* +
      \phi_\textit{sf}\widetilde{\phi}_\textit{nf}^* +
      \phi_\textit{sf}\phi_\textit{nf}^*
    \right]}{\left|\phi_\textit{nf}+\widetilde{\phi}_\textit{nf}\right|^2}.
\end{equation}
Here, hadronic $\phi$ and electromagnetic $\widetilde{\phi}$ parts of an amplitude are discriminated by tilde symbol.

For $t\!\to\!0$, there is a simple relation between {\em sf} and {\em nf} amplitudes\,\cite{Buttimore:1998rj}:
\begin{equation}
  \widetilde{\phi}_\textit{sf}/\widetilde{\phi}_\textit{nf}=\frac{\kappa_p}{2}\,\sqrt{-t}/m_p,\quad%
  \phi_\textit{sf}/\phi_\textit{nf}=\frac{r_5}{i+\rho}\,\sqrt{-t}/m_p,
  \label{eq:r5}
\end{equation}
where $m_p$ is a proton mass, $\kappa_p\!=\!\mu_p-1\!=\!1.793$ is anomalous magnetic moment of a proton, and complex $r_5\!=\!R_5\!+\!iI_5$, $|r_5|\!\sim\!0.02$\,\cite{Poblaguev:2019saw}, parameterize hadronic spin-flip amplitude \cite{ Buttimore:1998rj}. Omitting some small corrections\,\cite{Poblaguev:2019vho}, the analyzing power can be approximated\,\cite{Buttimore:1998rj} as
\begin{align}
  A_\text{N}&(t)=\frac{\sqrt{-t}}{m_p}\times \nonumber\\
  &\frac
  {\left[\kappa_p(1\!-\!\delta_C^{em}\rho)\!-\!2(I_5\!-\!\delta_C^{h}R_5)\right]{t_c/t}% 
    \!-\!2(R_5\!+\rho{I_5})}
       {\left({t_c/t}\right)^2 - 2\left(\rho+\delta_C\right){t_c/t}+1+\rho^2}%
       \!,
       \label{eq:AN}
\end{align}
where $\delta_C$ is given in Eq.\,(\ref{eq:dC}) while $\delta_C^{em}$ and $\delta_C^h$ are spin-flip phase shifts in  the $\widetilde{\phi}_\textit{sf}\phi_\textit{nf}^*$ and $\phi_\textit{sf}\widetilde{\phi}_\textit{nf}^*$ interference terms, respectively.

All recent experimental studies\,\cite{Poblaguev:2019saw,Okada:2005gu,Alekseev:2009zza,STAR:2012fiw} of the forward elastic proton-proton $A_\text{N}(t)$ had been done using Eq.\,(\ref{eq:AN}) and assuming  $\delta_C^{em}\!=\!\delta_C^h\!=\!\delta_C$.

\section{Spin-flip Coulomb phases in elastic $\boldsymbol{{p^\uparrow}\!\!\!{p}}$ analyzing power}

\begin{figure}
  \begin{center}
  \includegraphics[width=\columnwidth]{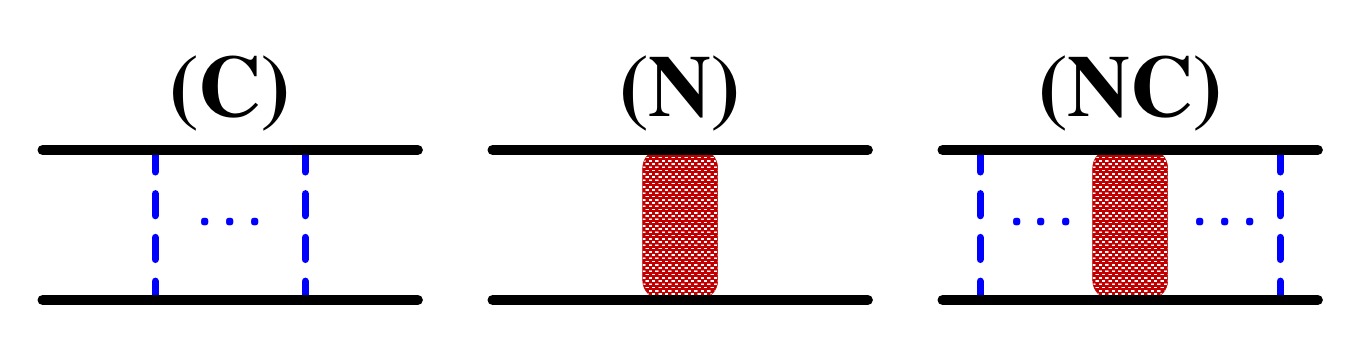}
  \end{center}
  \caption{ \label{fig:graphs}
    Three types of the elastic ${pp}$ scattering: (C) electromagnetic including multiphoton exchange, (N) bare hadronic, 
    and (NC) combined hadronic and electromagnetic.
  }
\end{figure}

Spin-flip phases, $\delta_C^{em}$ and $\delta_C^h$, can be evaluated in a simple way using expressions derived in Ref.\,\cite{Kopeliovich:2000ez} to study $\delta_C(t)$. To provide a framework for the calculations, some results of Ref.\,\cite{Kopeliovich:2000ez} are briefly overviewed in Sec.\,\ref{sec:KT}.

\subsection{Theoretical approach used\,\cite{Kopeliovich:2000ez} to calculate Coulomb corrections to the nonflip amplitudes
\label{sec:KT} }

Considering multiple photon exchange in the elastic $\textit{pp}$ scattering and neglecting the higher order corrections ${\cal O}(\alpha^3)$, the net long-range Coulomb (C) amplitude (see Fig.\ref{fig:graphs}) can be presented as\,\cite{Kopeliovich:2000ez}
\begin{align}
  f_C(q_T)  &= \frac{i}{2\pi}
  \int{d^2b\,e^{i\vec{q}_T\vec{b}}\left[1-e^{i\chi_C^\textit{nf}(b)}\right]},
  \label{eq:fC} \\
  &= \hat{f}_C(q_T) + \frac{i}{2\pi}\int{d^2b\,e^{i\vec{q}_T\vec{b}}\left[\chi_C(b)\right]^2/2}
\end{align}
where $q_T\!\approx\!\sqrt{-t}$ is transverse momentum and
the eikonal phase
\begin{equation}
  \chi_C^\textit{nf}(b)  =  \frac{1}{2\pi}
  \int{d^2q_T \hat{f}_C(q_T) e^{-i\vec{q}_T\vec{b}}}
\end{equation}
is a Fourier transform of the Coulomb part of the amplitude calculated in Born approximation\,\cite{Kopeliovich:2000ez}
\begin{equation}
  \hat{f}_C(q_T) = \frac{-2\alpha}{q_T^2+\lambda^2}\, e^{-\widetilde{B}q_T^2/2}.  
  \label{eq:fC_Born}
\end{equation}
Here, $\hat{f}_C$ is defined as sum of two nonflip helicity amplitudes
$\langle\!+\:\!\!+\!|\!+\:\!\!+\!\rangle$ and $\langle\!+\:\!\!-\!|\!+\:\!\!-\!\rangle$
\cite{Buttimore:1998rj}. A small photon mass $\lambda$ was included to Eq.\,(\ref{eq:fC_Born}) to keep the integrals finite.

The multiphoton exchange results in an acquired Coulomb phase $\Phi_C(q_T)$ 
\begin{equation}
  f_C(q_T) = \hat{f}_C(q_T)\,e^{i\Phi_C(q_T)}.
\end{equation}
Assuming $\Phi_C(q_T)\!\ll\!1$, one finds
\begin{align}
  \!\!\Phi_C(q_T) &=
  -i\left[f_C(q_T)/\hat{f}_C(q_T)-1\right]
  \\
  &=   \frac{1}{4\pi}\int{d^2q_1d^2q_2\delta(\vec{q}_T\!-\!\vec{q}_1\!-\!\vec{q}_2)%
    \frac{\hat{f}_C(q_1)\hat{f}_C(q_2)}{\hat{f}_C(q_T)}}\!,
  \label{eq:PhiC}
\end{align}

Similarly, to calculate the Coulomb corrections to the hadronic amplitude,
\begin{equation}
  \hat{f}_N(q_T) = \frac{(i+\rho)\sigma_\text{tot}}{4\pi}\,e^{-Bq_T^2/2},
\end{equation}
one can use the following relations,
\begin{align}
  f_\text{NC}(q_T) &= \frac{i}{2\pi}
  \int{d^2b\,e^{i\vec{q}_T\vec{b}}\gamma_N^\textit{nf}(b)e^{i\chi_C^\textit{nf}(b)}}
  \label{eq:fNC} \\
  &= \hat{f}_N(q_T) + \frac{i}{2\pi}
  \int{d^2b\,e^{i\vec{q}_T\vec{b}}\gamma_N^\textit{nf}\chi_C^\textit{nf}},
  \\
  \gamma_N^\textit{nf}(b) &=  \frac{-i}{2\pi}
  \int{d^2q_T e^{-i\vec{q}_T\vec{b}}\,\hat{f}_N(q_T)},
  \label{eq:gammaN} \\
   \Phi_\text{NC}(q_T) &= -\frac{\alpha}{\pi}\int{\frac{d^2q_1}{q_1^2+\lambda^2}}
  \nonumber \\
    &\times \exp{\left[
        -(B+\widetilde{B})q_1^2/2+B\vec{q}_1\vec{q_T}
        \right]}.
  \label{eq:PhiNC}
\end{align}

Equations (\ref{eq:PhiC}) and (\ref{eq:PhiNC}) were analytically integrated in Ref.\,\cite{Kopeliovich:2000ez}.
Both, $\Phi_C(q_T)$ and $\Phi_\text{NC}(q_T)$, contain the divergent term $\ln{q^2/\lambda^2}$ which, however, cancels in final expression (\ref{eq:dC}) for the Coulomb phase difference 
\begin{equation}
  \delta_C(t) = \Phi_C(t)-\Phi_\text{NC}(t).
\end{equation}

\subsection{Calculation of the spin-flip Coulomb phase}

To find the Coulomb corrected spin-flip amplitudes $f_C^\textit{sf}(q_T)$ and $f_\text{NC}^\textit{sf}(q_T)$, one can use the following eikonal phases\,\cite{Lapidus:1978xa}
\begin{equation}
  \chi_C^\textit{sf}(b) = \frac{1}{2\pi}
  \int{d^2q e^{-i\vec{q}\vec{b}}\times%
  \frac{\kappa_p}{2m_p}\,(\vec{n}\vec{q})\,\hat{f}_C(q)/2}
  \label{eq:chiC_sf}
\end{equation}
and
\begin{equation}
  \gamma_N^\textit{sf}(b) = \frac{-i}{2\pi}
  \int{d^2q e^{-i\vec{q}\vec{b}}\times%
  \frac{r_5}{(i+\rho)m_p}\,(\vec{n}\vec{q})\,\hat{f}_N(q)/2}
  \label{eq:gammaN_sf}
\end{equation}
respectively. Here,
$\hat{f}_C/2$ and $\hat{f}_N/2$ correspond to the nonflip amplitudes used in Eq.\,(\ref{eq:r5}), and the azimuthal dependence of the scattering is defined by 
$\vec{n}$, a unit vector orthogonal to the beam momentum and the proton spin.

Considering the spin flip amplitudes for $\vec{q}_T\!=\!\vec{n}q_T$, one can readily determine the spin-flip phase $\Phi_C^\textit{sf}(q_T)$ by including factor $2(\vec{q}_T\vec{q}_1)/q_T^2$ or $2(\vec{q}_T\vec{q}_2)/q_T^2$ to integral (\ref{eq:PhiC}).
Since  $\vec{q}_1\vec{q}_T\!+\!\vec{q}_2\vec{q}_T\!=\!q_T^2 $, we immediately find
\begin{equation}
  \Phi_C^\textit{sf}(q_T)\!=\!\Phi_C(q_T),
\end{equation}
which leads to
\begin{equation}
  \delta_C^\textit{em}(t) = \delta_C(t).
  \label{eq:deltaCem}
\end{equation}

To calculate $\Phi_\text{NC}^\textit{sf}(q_T)$, factor $(\vec{q}_2\vec{q}_T)/q_T^2 = 1 - (\vec{q}_1\vec{q}_T)/q_T^2$ should be applied in Eq.\,(\ref{eq:PhiNC}), which gives
\begin{align}
  \!\!\Phi_\text{NC}^\textit{sf}(q_T) &= \Phi_\text{NC}(q_T)+\frac{\alpha B}{B+\widetilde{B}}
  \times\Delta_\text{NC}^\textit{sf}(\eta), \label{eq:PhiNCsf}\\
  \Delta_\text{NC}^\textit{sf}(\eta) &= 
  \int_0^\infty{\frac{du}{\eta}e^{-u^2/4\eta}}%
  \int_{-\pi}^{\pi}{\frac{d\varphi}{2\pi}\cos{\varphi}\,e^{u\cos{\varphi}} }, \\
  \eta &= \frac{B}{B+\widetilde{B}}\times\frac{Bq_T^2}{2} \approx Bq_T^2/4.
\end{align}
Expanding
\begin{equation}
  e^{u\cos{\varphi}}\cos{\varphi} \to
  \sum_{k=0}^\infty{\frac{u^{k}}{k!}\cos^{k+1}{\varphi}}
\end{equation}
and using following definite integrals \cite{Gradshteyn:2014}
\begin{align}
  \int_{-\pi}^{\pi}{\cos^{2n+1}{x}\,dx} &= 0,
  \\
  \int_{-\pi}^{\pi}{\cos^{2n}{x}\,dx} &= \frac{\pi}{2^{2n-2}}\,\frac{(2n-1)!}{(n-1)!n!},
  \\
  \int_0^\infty{x^{2n+1}e^{-px^2}\,dx} &= \frac{n!}{2p^{n+1}},
\end{align}
one arrives to
\begin{equation}
  \Delta_\text{NC}^\textit{sf}(\eta) = \sum_{k=0}^\infty{\frac{\eta^k}{(k+1)!} }
  = \frac{e^\eta-1}{\eta}.
  \label{eq:DeltaNC}
\end{equation}
Thus,
\begin{equation}
  \delta_C^h(t) = \delta_C(t) - \frac{\alpha B}{B+\widetilde{B}}\,\frac{e^\eta-1}{\eta}.
  \label{eq:deltaCh}
\end{equation}

Evaluating $\delta_C^\textit{em}$ and $\delta_C^{h}$, we did not distinguish between nonflip $B$ and spin-flip $B_\textit{sf}$ hadronic slopes as well as between $\widetilde{B}$ [Eq.\,(\ref{eq:Bem})] and
\begin{equation}
  \widetilde{B}_\textit{sf} = \left(r_E^2+r_M^2\right)/3=12.25\pm0.25\,\text{GeV}^{-2},
\end{equation}
where $r_M=0.851\!\pm\!0.026\,\text{fm}$\,\cite{Lee:2015jqa} is rms magnetic radius of a proton.

For small $t$, i.e., omitting terms approaching zero if $t\to0$, one finds
\begin{align}
  \!\!\delta_C^\textit{em}(t,\widetilde{B}_\textit{sf},B) &= \delta_C(t)%
  + \alpha\ln{\frac{B+\widetilde{B}}{B+\widetilde{B}_\textit{sf}}}
  \\
  \delta_C^h(t,\widetilde{B},B_\textit{sf}) &= \delta_C(t)%
  - \frac{\alpha B_\textit{sf}}{B_\textit{sf}+\widetilde{B}}%
  + \alpha\ln{\frac{B+\widetilde{B}}{B_\textit{sf}+\widetilde{B}}}
\end{align}
Since $r_M\!=\!r_E$ within the current experimental accuracy of about 2\%, we cannot distinguish between $\widetilde{B}$ and $\widetilde{B}_\textit{sf}$. Also, there are arguments\,\cite{Kopeliovich:2021rdd} to assume that $B_\textit{sf}\approx B$. Although in an extreme case, $B_\textit{sf}\!=\!2B$ (e.g., considered in Ref.\,\cite{Cudell:2004ev} for proton-carbon scattering), $\delta_C^h\!-\!\delta_C$ can be increased by about a factor of 2, the effect will be invisible in the expression for analyzing power due to the strong suppression of term $\delta_C^h R_5$ in (\ref{eq:AN}) by a small value of $|R_5|\lesssim0.02$.

\subsection{The electromagnetic correction to $\boldsymbol r_5$}

In Ref.\,\cite{Kopeliovich:2021rdd}, it was pointed out that the hadronic spin-flip amplitude should also include the spin-flip photon exchange, i.e., one should replace
\begin{equation}
  \gamma_N^\textit{sf}(b) \to \gamma_N^\textit{sf}(b) + i\chi_C^\textit{sf}(b)\gamma_N^\textit{nf}(b)
  \label{eq:sfAll}.
\end{equation}
The effective spin-flip amplitude $r_5^\gamma$ can be related to the integral
\begin{equation}
  \frac{r_5^\gamma-r_5}{i+\rho}\,\frac{\vec{n}\vec{q}}{2m_p} =%
  \frac{i}{2\pi} \int{ d^2be^{i\vec{q}_T\vec{b}}\,\chi_C^\textit{sf}(b)\gamma_N^\textit{nf}(b)},
  \label{eq:Integral}
\end{equation}
which is similar to that of calculated in (\ref{eq:PhiNCsf})--(\ref{eq:DeltaNC}). Thus,
\begin{align}
  r_5^\gamma &=    r_5+i(i+\rho)\Delta_\gamma\approx r_5-\Delta_\gamma, \\
  \Delta_\gamma &= \frac{\kappa_p}{2}\frac{\alpha B}{(B+\widetilde{B}_\textit{sf})}\approx0.003.
 \label{eq:dr5}
\end{align}
For the modified $r_5^\gamma$, Coulomb phase $\Phi_\text{NC}^\textit{sf}(t)$ is the same as in Eq.\,(\ref{eq:PhiNCsf}).

Actually $r_5^\gamma$ had being determined in all previous measurements of $r_5$.

\section{Summary}

The technique developed in Ref.\,\cite{Kopeliovich:2000ez} was adapted for calculation of the Coulomb phase shifts in the spin-flip terms $\widetilde{\phi}_\textit{sf}\phi_\textit{nf}^*$\,(\ref{eq:deltaCem}) and $\phi_\textit{sf}\widetilde{\phi}_\textit{nf}^*$\,(\ref{eq:deltaCh}) of $A_\text{N}(t)$.

Small difference, $\delta_C^h\!-\!\delta_C\sim-\alpha/2$, was found for the forward elastic $\mathit{pp}$ scattering, $|t|\lesssim0.05\,\text{GeV}^2$ ($\eta<0.1$). Since $|r_5|\!\lesssim0.02$, such a discrepancy  can be neglected in Eq.\,(\ref{eq:AN}).

Thus, assuming spin-dependent measurements at high energies, we can agree with the approximation for the Coulomb phases  
\begin{equation}
  \delta_C^\textit{em} = \delta_C^\textit{h} = \delta_C =
   -\alpha\times 
   \left[
    \ln{\frac{(B\!+\!\widetilde{B})|t|}{2}} + \gamma\right]
\end{equation}
suggested in Ref.\,\cite{Buttimore:1998rj}.

Spin-flip photon exchange (\ref{eq:sfAll}) results in an effective correction $\Delta_\gamma\!\approx\!0.003$
to real part of the hadronic spin-flip parameter $r_5$ in Eq.\,(\ref{eq:AN}). The correction found is about triple of the experimental accuracy for $R_5$ in the HJET measurements \cite{Poblaguev:2019saw}.
Since $\Delta_\gamma$ is independent of the experimental data analysis, any already published experimental value $r_5^\gamma$ of the hadronic spin-flip amplitude, in particular given in\,\cite{Poblaguev:2019saw,STAR:2012fiw}, can be easily adjusted to a bare one,
\begin{equation}
  {\rm Re}\,r_5 = {\rm Re}\,r_5^\gamma + \frac{\kappa_p}{2}\frac{\alpha B}{B+\widetilde{B}_\textit{sf}}.
  \label{eq:r5Corr}
\end{equation}
This might be especially important for a study of $r_5(s)$ dependence on energy, e.g., in the Regge fit\,\cite{Poblaguev:2019saw,Kopeliovich:2021rdd}.

To illustrate a possible effect of corrections (\ref{eq:r5Corr}), the Regge fit\,\cite{Poblaguev:2019saw} of the HJET values of $r_5$ (for $\sqrt{s}\!=\!13.76~\text{and}~21.92\,\text{GeV}$),
\begin{equation}
  \sigma_\text{tot}(s)\times r_5(s) =
  \sum_{{\cal{}R}=P,R^\pm}{f_5^{\cal{}R}\,{\cal{}R}(s)},
  \label{eq:r5Fit}
\end{equation}
was revisited.
In Ref.\,\cite{Poblaguev:2019saw}, the spin-flip Reggeon, $R^+(s)$ and $R^-(s)$, and Pomeron (in a Froissaron parametrization), $P(s)$, functions were approximated by the nonflip ones\,\cite{Fagundes:2017iwb}.
After applying corrections (\ref{eq:r5Corr}), the fit $\chi^2\!=\!2.2$ (NDF=1) was improved to $\chi^2\!=\!0.9$ and the central values of couplings $f_5^{\cal{}R}$ were shifted to
\begin{align}
  \phantom{-}f_5^{P^{\phantom+}}\!\!\!=\! 0.045 \!\pm\! 0.002_\text{stat} \!\pm\! 0.003_\text{syst}
  &\,\to\, 0.053,    \label{eq:f5P}\\
  -f_5^{R^+}       \!\!\!=\!  0.032\!\pm\! 0.007_\text{stat} \!\pm\! 0.014_\text{syst}
  &\,\to\, 0.069,  \label{eq:f5R+}\\
  \phantom{-}f_5^{R^-}       \!\!\!=\! 0.622\!\pm\! 0.023_\text{stat} \!\pm\! 0.024_\text{syst}
    &\,\to\, 0.654.  \label{eq:f5R-} 
\end{align}
A thorough refit of the $r_5$ measurements, including the STAR $\sqrt{s}\!=\!200\,\text{GeV}$ result\,\cite{STAR:2012fiw}, will be done elsewhere.

It is interesting to note that an absorptive correction, 
\begin{equation}
  a_\textit{sf} = \alpha B / (B+\widetilde{B}_\textit{sf}),
  \label{eq:sfAbs}
\end{equation}
to the electromagnetic spin-flip form factor, ${\cal F}_{pp}^\textit{sf}(t) \to {\cal F}_{pp}^\textit{sf}(t)\times%
\left( 1 + a_\textit{sf}\,t/t_c\right)$, results\,\cite{Poblaguev:2019vho} in the same effective alteration of $r_5$ as shown in Eq.\,(\ref{eq:r5Corr}).

In Ref.\,\cite{Kopeliovich:2021rdd}, to calculate absorptive corrections, graphs in Fig.\,\ref{fig:graphs} were regrouped. In this approach, a spin-flip photon contribution [Eq.\,(\ref{eq:sfAll}] to hadronic spin-flip amplitude $f_N^\textit{sf}$ was not considered. Nonetheless, the term $\chi_C^\textit{sf}(b)\gamma_N^\textit{nf}(b)$ appeared as an absorptive correction to electromagnetic spin-flip amplitude $f_C^\textit{sf}(q_T)$. The corresponding equation was integrated numerically in \cite{Kopeliovich:2021rdd}. However, using result of calculation (\ref{eq:Integral}), one can readily find absorption parameter $a_\textit{sf}$ to be in exact agreement with Eq.\,(\ref{eq:sfAbs}). Thus, we come to a conclusion\,\cite{Kopeliovich:2021rdd} that Coulomb corrections to the hadronic spin-flip amplitude (\ref{eq:r5Corr}) and hadronic (absorption) correction to the spin-flip Coulomb amplitude (\ref{eq:sfAbs}) are two equivalent descriptions of the same final-state electromagnetic interaction of polarized protons.

\section*{Acknowledgments}

The author is thankful to Nigel Buttimore, Boris Kopeliovich, and Michal Krelina for informative discussions and acknowledges support from the Office of Nuclear Physics in the Office of Science of the US Department of Energy. This work is authored by employees of Brookhaven Science Associates, LLC under Contract No. DE-SC0012704 with the U.S. Department of Energy.

\bibliographystyle{apsrev4-2}
%\bibliography{v4.2_CoulombPhases_PRD.bib}
\input{v4.2_CoulombPhases_arXiv.bbl}

\end{document}